\documentclass{ws-procs9x6}

\newcommand{\be}{\begin{equation}}
\newcommand{\ee}{\end{equation}}

\newcommand{\bea}{\begin{eqnarray}}
\newcommand{\eea}{\end{eqnarray}} 

\begin{document}

\title{Prethermalisation and \\ the build-up of the Higgs effect}

\author{D. SEXTY and A. PATK\'OS\footnote{\uppercase{W}ork 
supported by grant \uppercase{T}-037689 of the \uppercase{H}ungarian 
\uppercase{N}ational \uppercase{S}cience \uppercase{F}oundation.}}

\address{Department of Atomic Physics \\
E{\"o}tv{\"o}s University\\ 
H-1117 P\'azm\'any P{\'e}ter s{\'e}t\'any 1/A\\ 
E-mail: denes@achilles.elte.hu, patkos@ludens.elte.hu}

\maketitle

\abstracts{Real time field excitations in the broken symmetry phase
of the classical abelian Gauge+Higgs model are studied
numerically in the
unitary gauge, for systems starting from the unstable
maximum of the Higgs potential.}

\section{Introduction}

Tracking the transmutation of the angular component of a complex Higgs
field into the longitudinal polarisation state of the gauge field
during the termination of inflation in hybrid models
 might reveal interesting details of the real time
Higgs effect and of  the electroweak dynamics \cite{smit04}.  The
excitation rate of the different polarisation
states might be different as well as the thermal relaxation rates
\cite{skullerud03}. The production of gauged cosmic
strings is another important aspect of this process\cite{hindmarsh01}. Our
numerical study concentrates on the
non-equilibrium phase transition aspects.

\section{Partial Pressures and Energy Densities of the Model}

The main observables studied in the present
 investigation of  the classical abelian Gauge+Higgs model,
\be
L=-\frac{1}{4}F_{\mu\nu}F^{\mu\nu}+\frac{1}{2}D_\mu\Phi(D^\mu\Phi)^*-V(\Phi),
\quad V(\Phi)=\frac{1}{2}m^2\Phi^2+\frac{\lambda}{24}\Phi^4
\label{Lagrangian}
\ee
are the partial energy densities and pressures:
\bea
\epsilon&=&\epsilon_\rho+\epsilon_T+\epsilon_L,\quad 
p=p_\rho+p_T+p_L,\nonumber\\
\epsilon_\rho&=&\frac{1}{2}\Pi_\rho^2+\frac{1}{2}(\nabla\rho)^2+V(\rho),
\quad p_\rho=\frac{1}{2}\Pi_\rho^2-\frac{1}{6}(\nabla\rho)^2-V(\rho),
\nonumber\\ 
\epsilon_T&=&\frac{1}{2}[{\bf \Pi}_T^2+(\nabla\times{\bf
    A}_T)^2+e^2\rho^2{\bf A}_T^2],\nonumber\\
p_T&=&\frac{1}{6}[{\bf \Pi}_T^2+(\nabla\times{\bf
    A}_T)^2-e^2\rho^2{\bf A}_T^2],\nonumber\\
\epsilon_L&=&\frac{1}{2}\left[{\bf \Pi}_L^2+e^2\rho^2\left({\bf
    A}_L^2+\frac{1}{(e^2\rho^2)^2}(\nabla\Pi_L)^2\right)\right],\nonumber\\
p_L&=&\frac{1}{6}[{\bf \Pi}_L^2-e^2\rho^2{\bf A}_L^2]+
\frac{1}{2}\frac{1}{e^2\rho^2}(\nabla{\bf\Pi}_L)^2.
\label{energy-pressure}
\eea
Index $T$ refers to the transversal, 
$L$ to the longitudinal part of the gauge field $\bf A$. The
expressions are valid in the unitary gauge: $\rho=|\Phi|$. $A_0$ was
eliminated with the Gauss constraint.
It was checked numerically that in equilibrium
 $\epsilon_\rho:\epsilon_L:\epsilon_T=1:1:2$ within
statistical fluctuations \cite{sexty04}.

The equations of motion derived from (\ref{Lagrangian}) were solved in the
$A_0=0$ gauge and the solutions were transformed to the unitary gauge for
measurements. Initially $\rho({\bf x},t=0)=0$ was put uniformly.
Inhomogenuous modes beyond the region
of spinodal instability (up to $|{\bf k}|\leq 5|m|)$ were filled with low
amplitude white noise. The spatial lattice spacing was chosen to
$a_s=0.35|m|^{-1}$ and the time step $\delta t=0.04 |m|^{-1}$.

\section{Spectral Analysis of the Process of Excitation}
 
 Spectral representation of the energy densities $\epsilon ({\bf k})$
   (pressures $p({\bf k})$) were defined using the Fourier transform of
 the square root of various pieces of the local
  energy densities (pressures) and taking their absolute squares.

\begin{figure}[ht]
\centerline{\epsfxsize=5cm\epsfbox{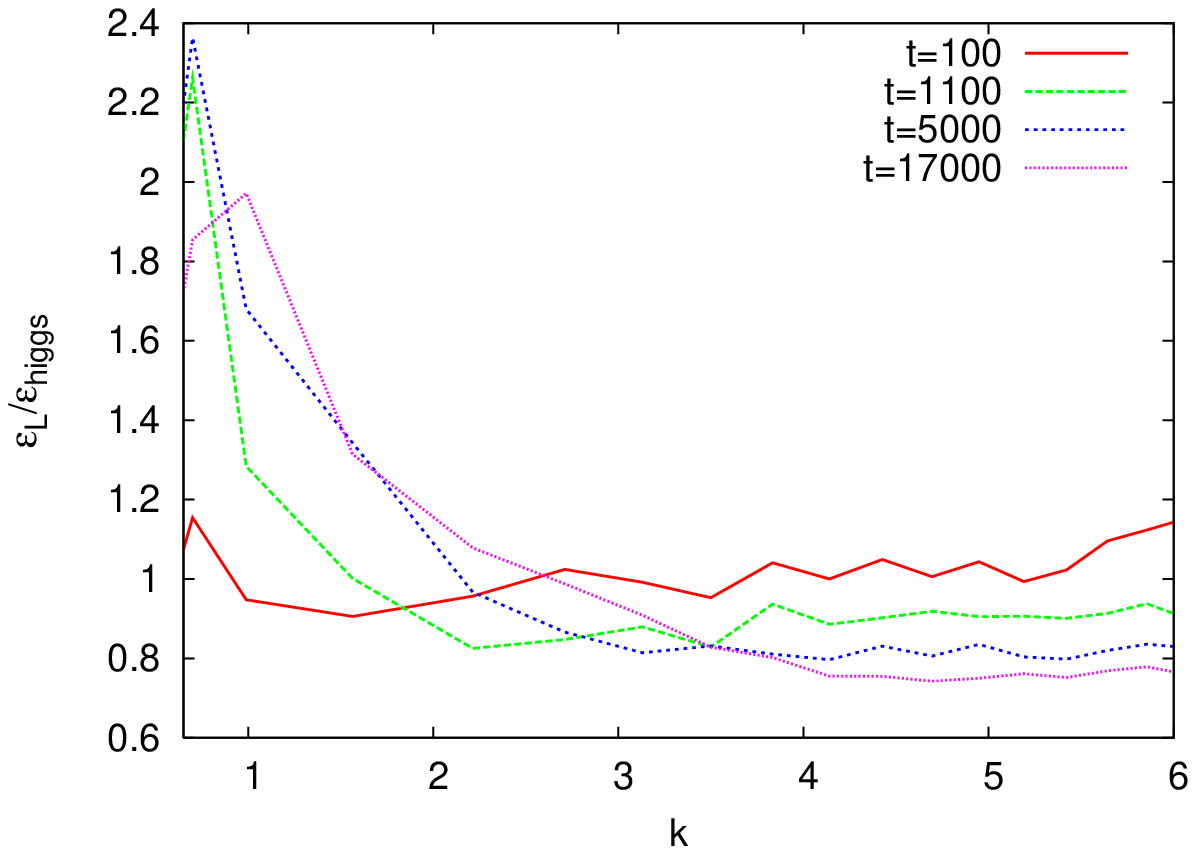}
\epsfxsize=5cm\epsfbox{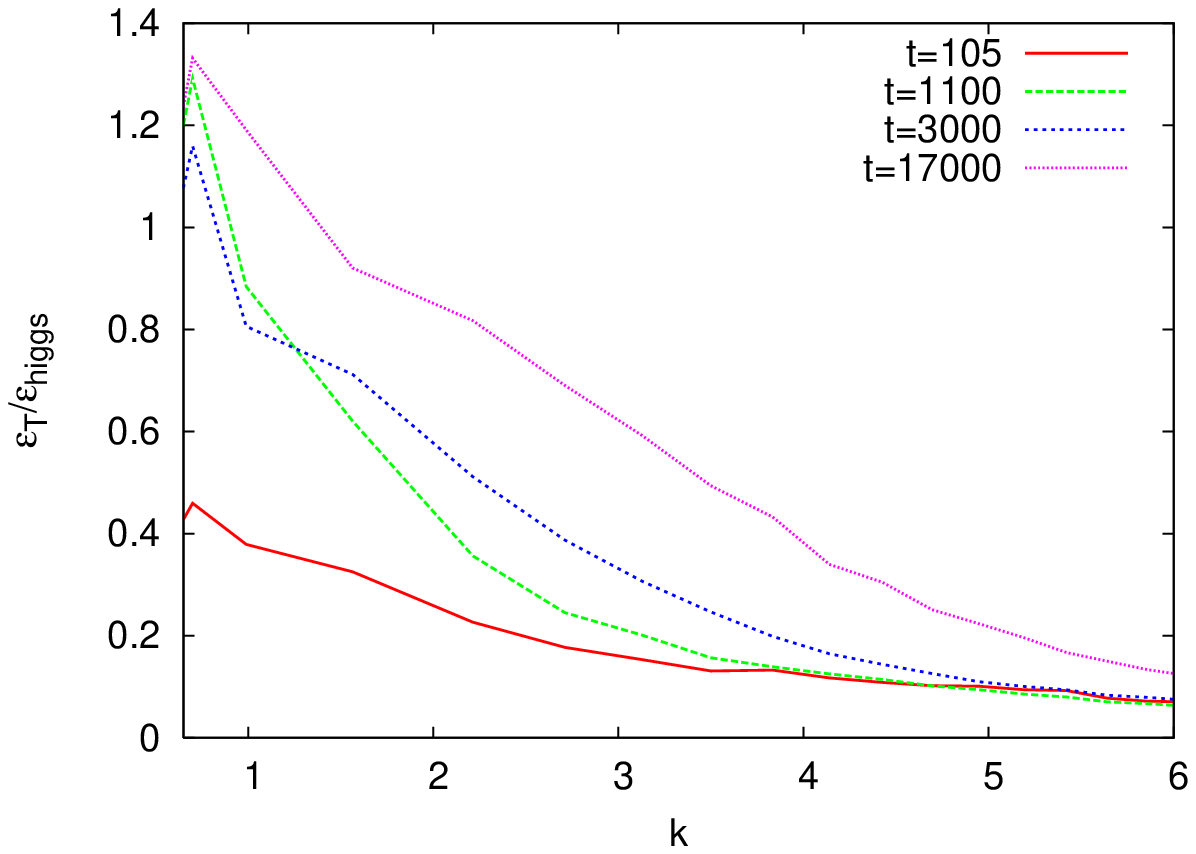}}   
\caption{ Evolution of the longitudinal (left) and
  transversal (right) energy excitations
relative to the ${\bf k}\neq 0$ Higgs modes.
 \label{epere}}
\end{figure}

In {\bf Fig.\ref{epere}} the degrees of excitation $\epsilon_T({\bf
   k})/\epsilon_\rho({\bf k})$ and $\epsilon_L({\bf
   k})/\epsilon_\rho({\bf k})$ are displayed as functions of $|{\bf
   k}|$ at different times.  The
   low frequency part of $\epsilon_L({\bf k})$ develops an
   over-excited peak, while $\epsilon_T({\bf k})$ is excited weakly
  at early times. Their slow evolution towards
   equipartition defines an extremely long thermalisation time-scale 
($\tau>10^5|m|^{-1}$).

\section{Signals for Prethermalisation}

\begin{figure}[ht]
\centerline{\epsfxsize=5cm\epsfbox{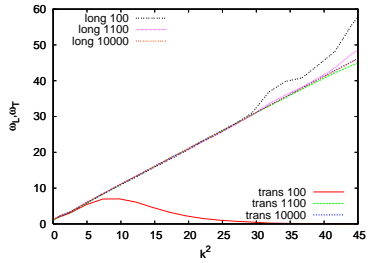}}   
\caption{ Evolution of the longitudinal and transversal dispersion relations
 \label{disprel}}
\end{figure}

 The dispersion relations $\omega^2({\bf k})$
of the modes $(\rho, {\bf A}_T)$
are obtained as $|\Pi_\rho({\bf k})|^2/|\rho({\bf k})|^2$
and $|\Pi_T({\bf k})|^2/|{\bf A}_T({\bf k})|^2$. For the longitudinal
mode inspection of $\epsilon_L$ in (\ref{energy-pressure}) suggests 
$
\omega_L^2({\bf k})\equiv |[e^2\rho^2{\bf A}_L]({\bf
  k})|^2/|\Pi_L({\bf k})|^2.
$
 The masses of the longitudinal and transversal modes
 become degenerate early when calculated from modes  
$1\leq |{\bf k}|\leq 5$; evolutionary effects are seen only at high
 $k$  (see {\bf Fig. \ref{disprel}}).

Spectral equations of state for the
different fields can be defined assuming the mode-by-mode equality of
the kinetic and potential spectral energy densities.
 Using this in (\ref{energy-pressure}) one arrives at the
following equation for ${\bf A}_T$:
\be
w_T({\bf k})\equiv p_T({\bf k})/
{\epsilon}_T({\bf k})={\bf k}^2|{\bf A}_T({\bf k})|^2/3
|\Pi_T({\bf k})|^2={\bf k}^2/3\omega_T^2({\bf k}).
\label{eqstat}
\ee
Similar relation holds for $\rho$. Using $\omega_L^2({\bf k})$ for the
longitudinal mode the same formula appears on the right end of
the chain if the definition $w_L({\bf
  k})\equiv[e^2\rho^2p_L]({\bf k})/[e^2\rho^2\epsilon_L]({\bf k})$ is applied.
The expected functional form (with newly fitted squared mass values) 
was compared with the measured $w_{T,L}(|{\bf k}|)$.
The modes filled
initially almost instantly obey the expected form (\ref{eqstat}). 
Higher $|{\bf k}|$ modes are gradually filled and 
$w({\bf k})$ ''climbs up'' to the stable free particle behavior. These prompt
features illustrate the phenomenon of ''prethermalisation''
\cite{borsanyi04} in a gauge system.

\section{Gauge-Higgs Cross-Correlations}

The intiuitive quasi-particle picture in the unitary gauge conjectures
   that at low enough temperature and at moderate couplings the
   statistically independent field variables are just $\rho, {\bf
   A}_L, {\bf A}_T$.  In the
    analysis of the equations of state above we avoided to rely on 
the statistical independence of
   these three variables which we are going to test next.
 
\begin{figure}[ht]
\centerline{\epsfxsize=5cm\epsfbox{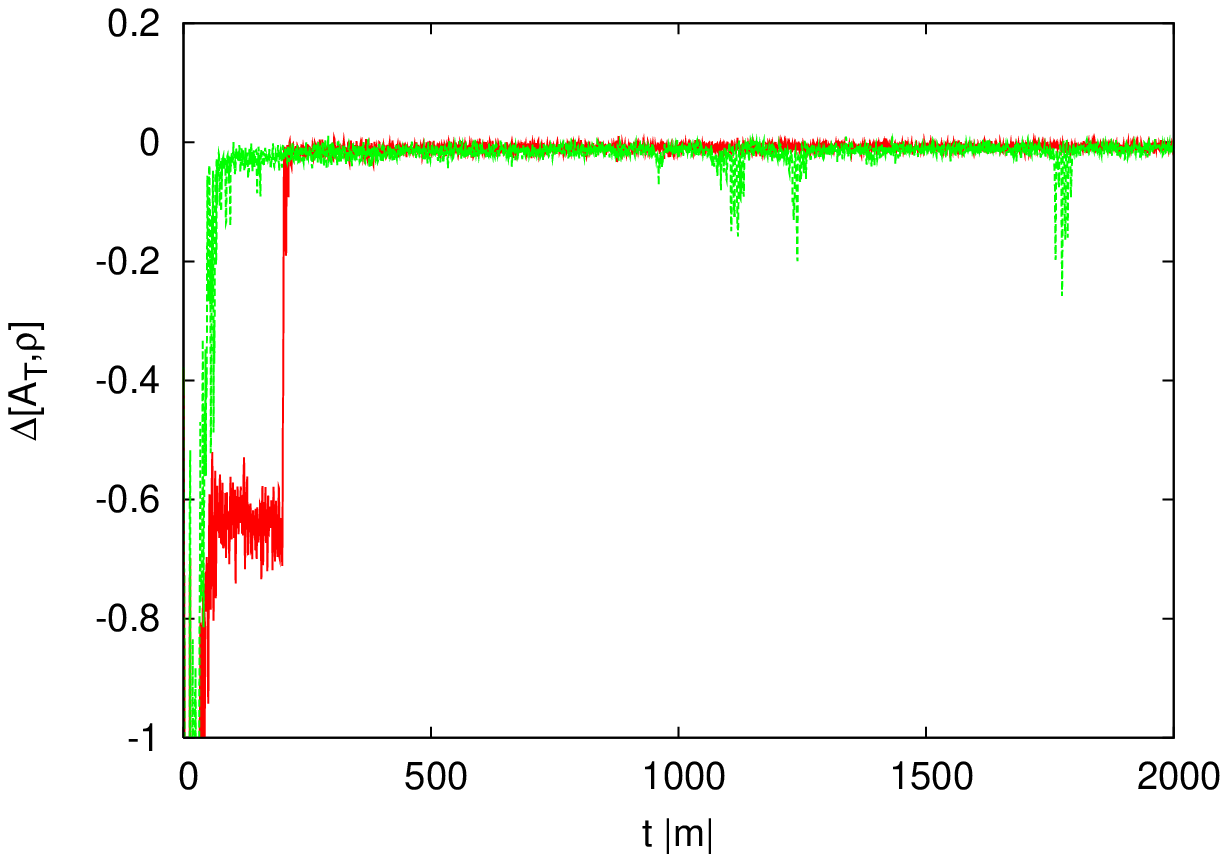}\epsfxsize=5cm\epsfbox{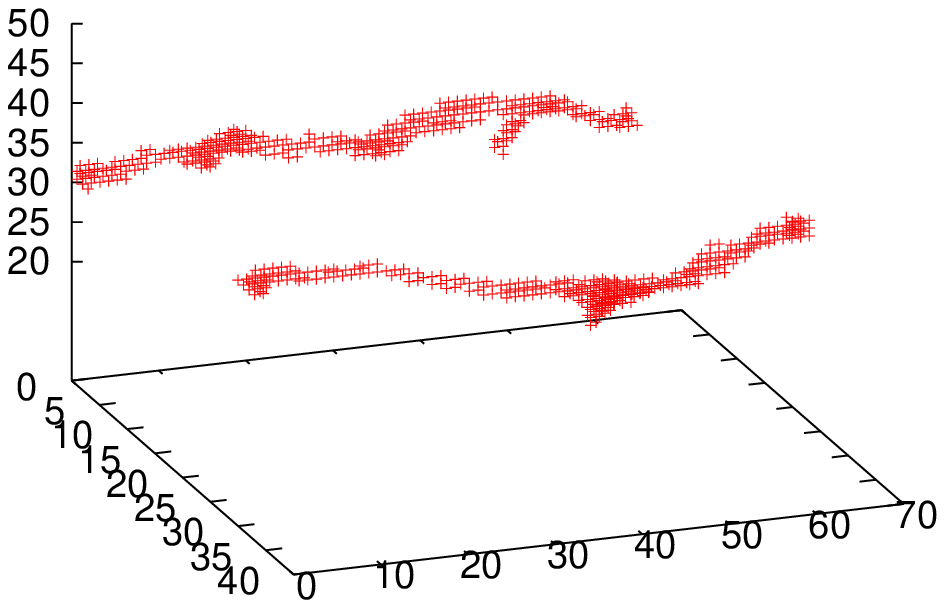}}   
\caption{ Non-smooth ${\bf A}_T-\rho$ correlation histories, and
  the corresponding  vortex pair.
 \label{transd}}
\end{figure}

{\bf The transverse polarisation.} The correlation
coefficient between the quadratic spatial averages of 
$\rho$ and ${\bf A}_T$ is defined as
\be
\displaystyle
\Delta[{\bf A}_T,\rho]\equiv
\left(\overline{\rho^2({\bf x},t){\bf A}_T^2({\bf x},t)}-
\overline{\rho^2({\bf x},t)}~\overline{{\bf A}_T^2({\bf x},t)}\right)/
\overline{\rho^2({\bf x},t){\bf A}_T^2({\bf x},t)}.
\label{corr-trans}
\ee 
In {\bf Fig.\ref{transd}a} the time evolution of this quantity is
displayed in two characteristic
runs. In the first a large negative value is reached almost
instantly after the Higgs-field rolls down. After a longer
time interval $\Delta[{\bf A}_T,\rho]$ suddenly jumps to a value compatible
with zero. In the other run one can observe
negative ''needles'' occuring on the background of near-zero
fluctuations. 

A large negative value of the correlation coefficient
(\ref{corr-trans}) represents
a very sensitive indicator for the presence of relativistic
Abrikosov-strings.  The location of the points where $\rho/|m|<
0.3$ displays  a vortex network very nicely (see {\bf Fig. \ref{transd}b}), 
but $\Delta\approx 0$ excludes the presence of vortices. 

{\bf The longitudinal polarisation.} The quantity $\Delta[{\bf A}_L,\rho]$
 always displays instantly after
the roll-down values significantly different from zero. and shows
 only a
very mild time variation. It was checked that a non-zero value
is present also in equilibrium \cite{sexty04}. 
This correlation coeficient linearly increases
with the temperature. These observations point to the fact that the
true quasiparticle field is
a composite of $\rho$ and ${\bf A}_L$. This is not a very great
surprise for relatively strongly coupled systems, still it should be
confronted with the fact that (without vortices) ${\bf A}_T({\bf
 k},t)$ and  $\rho ({\bf k},t)$ perform statistically independent and Gaussian 
 small oscillations. 

\begin{figure}[ht]
\centerline{\epsfxsize=5cm\epsfbox{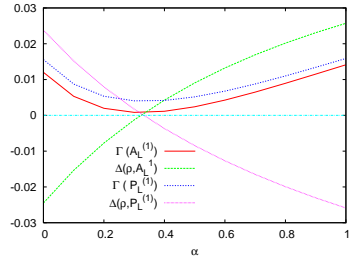}}   
\caption{ Correlations testing independence and Gaussianity of (\ref{nonlin})
in function of $\alpha$
 \label{alfa}}
\end{figure}

We have tested several trial composite fields at equilibrium \cite{sexty04}. 
In {\bf Fig. \ref{alfa}} we show
the variation of $\Delta[P_L,\rho],\Delta[Q_L,\rho]$ and of
$\Gamma[P_L],\Gamma[Q_L]$ defined by the ratio
$
\Gamma[Q_L]\equiv
\left({\overline{\left(Q_L^2({\bf x},t)\right)^2}-3\left(
\overline{Q_L^2({\bf x},t)}\right)^2}\right)/
\overline{\left(Q_L^2({\bf x},t)\right)^2}
$
as a function of $\alpha$ which characterizes the
compositeness of the conjugate variables
\be
Q_L({\bf x},t)=(1+\alpha\rho^2({\bf x},t)){\bf A}_L({\bf x},t),
\quad P_L({\bf x},t)=\Pi_L({\bf x},t)/(1+\alpha\rho^2({\bf x},t)).
\label{nonlin}
\ee
The coefficients $\Gamma$ signal deviations from Gaussianity. 
The nice surprise is that in equilibrium there exists a single optimal choice
$\alpha_{opt}$ where
$\Delta[P_L,\rho], \Delta[Q_L,\rho]$ vanish and both $\Gamma$'s are
minimal.
It turns out, however, that during the non-equilibrium phase
transition no such $\alpha_{opt}$ appears to exist, the ''longitudinal
quasiparticle'' coordinate presumably emerges only on the
thermalisation scale.

In conclusion, we found that in the abelian Higgs model
an early quasi-particle characterisation works well for the dispersion
relations and the equations of state just  after the
symmetry breaking is completed. The  non-linear quasi-particle field
containing the longitudinal vector component
 builds up much more slowly in the process of complete thermalisation.

\end{document}